\documentclass[11pt]{amsart}
\usepackage{amsmath,amssymb} 
\usepackage{graphicx}
\usepackage[utf8]{inputenc} 
\usepackage[font=small,labelfont=bf]{caption}

\begin{document}

\newtheorem{thm}{Theorem}[section]
\newtheorem{lem}[thm]{Lemma}
\newtheorem{prop}[thm]{Proposition}
\newtheorem{coro}[thm]{Corollary}
\newtheorem{defn}[thm]{Definition}
\newtheorem*{remark}{Remark}

\numberwithin{equation}{section}

\newcommand{\Z}{{\mathbb Z}} 
\newcommand{\Q}{{\mathbb Q}}
\newcommand{\PP}{{\mathbb P}}
\newcommand{\R}{{\mathbb R}}
\newcommand{\C}{{\mathbb C}}
\newcommand{\N}{{\mathbb N}}
\newcommand{\FF}{{\mathbb F}}
\newcommand{\T}{{\mathbb T}}
\newcommand{\fq}{\mathbb{F}_q}

\newcommand{\fixmehidden}[1]{}

\def\scrA{{\mathcal A}}
\def\cB{{\mathcal B}}
\def\Eps{{\mathcal E}}
\def\cI{{\mathcal I}}
\def\scrD{{\mathcal D}}
\def\cF{{\mathcal F}}
\def\cL{{\mathcal L}}
\def\cM{{\mathcal M}}
\def\cN{{\mathcal N}}
\def\cP{{\mathcal P}}
\def\scrR{{\mathcal R}}
\def\scrS{{\mathcal S}}

\newcommand{\rmk}[1]{\footnote{{\bf Comment:} #1}}

\renewcommand{\mod}{\;\operatorname{mod}}
\newcommand{\ord}{\operatorname{ord}}
\newcommand{\TT}{\mathbb{T}}
\renewcommand{\i}{{\mathrm{i}}}
\renewcommand{\d}{{\mathrm{d}}}
\renewcommand{\^}{\widehat}
\newcommand{\HH}{\mathbb H}
\newcommand{\Vol}{\operatorname{vol}}
\newcommand{\area}{\operatorname{area}}
\newcommand{\tr}{\operatorname{tr}}
\newcommand{\norm}{\mathcal N} 
\newcommand{\intinf}{\int_{-\infty}^\infty}
\newcommand{\ave}[1]{\left\langle#1\right\rangle} 
\newcommand{\E}{\mathbb E}
\newcommand{\Var}{\operatorname{Var}}
\newcommand{\Cov}{\operatorname{Cov}}
\newcommand{\Prob}{\operatorname{Prob}}
\newcommand{\sym}{\operatorname{Sym}}
\newcommand{\disc}{\operatorname{disc}}
\newcommand{\CA}{{\mathcal C}_A}
\newcommand{\cond}{\operatorname{cond}} 
\newcommand{\lcm}{\operatorname{lcm}}
\newcommand{\Kl}{\operatorname{Kl}} 
\newcommand{\leg}[2]{\left( \frac{#1}{#2} \right)}  
\newcommand{\id}{\operatorname{id}}
\newcommand{\beq}{\begin{equation}}
\newcommand{\eeq}{\end{equation}}
\newcommand{\bsp}{\begin{split}}
\newcommand{\esp}{\end{split}}
\newcommand{\bra}{\left\langle}
\newcommand{\ket}{\right\rangle}
\newcommand{\diam}{\operatorname{diam}}
\newcommand{\supp}{\operatorname{supp}}
\newcommand{\dist}{\operatorname{dist}}
\newcommand{\sgn}{\operatorname{sgn}}
\newcommand{\inte}{\operatorname{int}}
\newcommand{\Spec}{\operatorname{Spec}}
\newcommand{\ddiv}{\operatorname{div}}
\newcommand{\sumstar}{\sideset \and^{*} \to \sum}

\newcommand{\LL}{\mathcal L} 
\newcommand{\sumf}{\sum^\flat}
\newcommand{\Hgev}{\mathcal H_{2g+2,q}}
\newcommand{\USp}{\operatorname{USp}}
\newcommand{\conv}{*}
\newcommand{\CF}{c_0} 
\newcommand{\kerp}{\mathcal K}

\newcommand{\gp}{\operatorname{gp}}
\newcommand{\Area}{\operatorname{Area}}

\title[Multifractality for intermediate quantum systems]{Multifractality for \\ intermediate quantum systems}

\author{Henrik Uebersch\"ar}
\address{Sorbonne Universit\'e, Universit\'e Paris Cit\'e, CNRS, IMJ-PRG, F-75006 Paris, France.}
\email{henrik.ueberschar@imj-prg.fr}
\date{\today}

\begin{abstract}
While quantum multifractality has been widely studied in the physics literature and is by now well understood from the point of view of physics, there is little work on this subject in the mathematical literature. I will report on a proof of multifractal scaling laws for  arithmetic \u{S}eba billiards. I will explain the mathematical approach to defining the Renyi entropy associated with a sequence of eigenfunctions and sketch how arithmetic methods permit us to obtain a precise asymptotic in the semiclassical regime and how this allows us to compute the fractal exponents explicitly. Moreover, I will discuss how the symmetry relation for the fractal exponent is related to the functional equation of certain zeta functions. 
\end{abstract}
 
\maketitle 
 
\section{Introduction}
Many dynamical systems are in a state of transition between two regimes. In models of the brain,
such as neural networks, the firing patterns of neurons may undergo a transition from isolated firing
to avalanches of firing neurons. In the quantum physics of disordered electronic systems the system
may be in an insulating or a conducting phase. The former phase corresponds to electronic states
which are localized (no transport), whereas the latter phase corresponds to extended states (diffusive
dynamics). The study of phase transitions and, in particular, the critical states at the transition
between these different regimes, is central to understanding important phenomena such as the
functioning of our brain or the properties of semi-conducting materials.

One of the key features of systems in a critical state is that they often display a self-similarity in a
certain scaling regime which is so complex that it cannot be captured by a single fractal exponent
but only by a continuous spectrum of fractal exponents. This phenomenon is known as
multifractality.

Multifractality in quantum systems has been studied in the physics literature since the 1980s and
has become an extremely active field in theoretical and experimental physics \cite{AKL86, SchGr91,
KMu97, MaGGiGe10, BoGi11, AtBo12, BHK19, BiGGeGi20}. However, the abundance of results
in the physics literature is in stark contrast with a glaring absence of rigorous mathematical
results. One of the key difficulties in obtaining a mathematical proof is to formulate the problem in
a concise mathematical way and to then develop the mathematical methods which permit its
resolution.

In joint work with Keating, we recently proved the existence
of multifractal eigenfunctions for arithmetic \u{S}eba billiards \cite{KeU21} as well as quantum star graphs \cite{KeU22}. 
The key idea which permitted this advance was an approach to associate a quantity, known as Renyi's entropy -- in some sense a
generalization of Shannon's entropy -- with each eigenfunction. We were able to obtain asymptotic
estimates of the Renyi entropy along a typical sequence of eigenfunctions. This permitted the
derivation of explicit formulae for the fractal exponents and led to the derivation of a multifractal
scaling law for this system.

Multifractal self-similarity typically emerges at the transition between two physical regimes.
Examples of such intermediate quantum systems are disordered systems at the Anderson or
Quantum Hall transitions from a localized to a delocalized phase \cite{AKL86, SchGr91}. In the field of
quantum chaos, pseudo-integrable systems \cite{RiBe81} are intermediate between integrability and
chaos in the sense that their dynamics in phase space is not constrained to tori but rather to handled
spheres (e. g. rational polygonal billiards). One often includes in this class toy models of pseudointegrable
dynamics such as parabolic automorphisms of the torus \cite{MRu00}, quantum star graphs
\cite{BeK99, BeBoK01, KeMW03, BeKW04} and \u{S}eba billiards (rectangular billiards with a Dirac
delta potential) \cite{Se90}.

The morphology of eigenfunctions with multifractal self-similar structure is far more complex than
being purely localized or delocalized. Numerical and experimental studies of a large class of
quantum systems have resulted in numerous conjectures in the physics literature \cite{MaGGiGe10,
AtBo12, BHK19, BiGGeGi20} such as predictions of a symmetry relation for the fractal exponents
$D_q$ around the critical value $q=1/4$.

\section{The gap between localization and delocalization}
Much of the mathematical literature on quantum chaos over the past 40 years has focused on the
classification of limit measures which arise in the high frequency limit from eigenfunctions of
quantized chaotic systems. One of the key results of the field is the Quantum Ergodicity Theorem
which states that on a Riemannian manifold without boundary, whose geodesic flow is ergodic
with respect to Liouville measure, a typical sequence of eigenfunctions gives rise to Liouville
measure as the only semiclassical defect measure along this sequence.

Quantum Ergodicity (QE) was first proved in the 1980s by Zelditch and Colin de Verdi\`ere \cite{CdV85,
Z87} who completed the earlier work of Snirelman \cite{Sn74}. QE was later generalized to manifolds
with boundary by G\'erard-Leichtnam \cite{GeLe93} and Zelditch-Zworski \cite{ZZ96}.
The Quantum Unique Ergodicity (QUE) Conjecture put forward by Rudnick and Sarnak in 1994
\cite{RuSa94} asserts that the only such measure should be the Liouville measure. Lindenstrauss \cite{L06}
proved this conjecture in 2006 for arithmetic hyperbolic surfaces and was awarded the Fields Medal
for his work. Moreover, Anantharaman \cite{A08} ruled out localization on points or geodesic segments
for Anosov manifolds. De Bi\`evre-Faure-Nonnenmacher \cite{FNdB03} demonstrated the existence of partially
localized limit measures for the eigenstates of quantized hyperbolic automorphisms of tori with
minimal periods.

While rigorous mathematical work has largely focused on the proof of localization and
delocalization results for the probability densities which arise from quantum eigenfunctions
(Q(U)E, Scarring, Anderson localization), the key feature of intermediate quantum systems is
the multifractal self-similarity of their eigefunctions. This feature today remains
poorly understood from a mathematical point of view.

\section{Multifractality for Quantum Billiards}
Consider the Dirichlet problem for the positive Laplacian $-\Delta=-\partial_x^2-\partial_y^2$ on a compact domain $D\subset\R^2$ with piece-wise smooth boundary.  We have discrete spectrum accumulating at infinity associated with eigenfunctions $\psi_j$: 
\beq
(\Delta+\lambda_j)\psi_j=0, \quad \psi_j|_{\partial D}=0, 
\eeq
where
$$0=\lambda_0<\lambda_1\leq\cdots\leq \lambda_j\leq \cdots \to +\infty.$$

Our goal is to prove a multifractal scaling law for a subsequence of eigenfunctions $\{\lambda_{j_k}\}_{k=0}^\infty$, as $\lambda_{j_k}\to+\infty$. The general idea is to embed the domain in a rectangle and expand with respect to an eigenbasis of complex exponentials. A key point is that the scaling law should be independent of rotations and scaling of the rectangle in which the domain is embedded. The scaling parameter will then arise from the number of $O(1)$ contributions in this expansion, as the eigenvalue tends to infinity.

We will illustrate this in detail in the case of toral Schr\"odinger operators. Let $\T^d=2\pi\R^d/\Z^d$ and $V\in C^0(\T^d)$. We consider $L^2$-normalized solutions of the stationary Schr\"odinger equation on $\T^d$:
\beq
(-\Delta+V)\psi_\lambda=\lambda\psi_\lambda, \quad \|\psi_\lambda\|_{L^2(\T^d)}=1
\eeq
We can expand the eigenfunctions into Fourier series
\beq
\psi_\lambda(x)=\frac{1}{2\pi}\sum_{\xi\in \Z^d}\hat{\psi}_\lambda(\xi)e^{i\xi\cdot x}.
\eeq
By Parseval's identity, we obtain a discrete probability measure on $\Z^d$: $$\mu_\lambda(\xi):=|\hat{\psi}_\lambda(\xi)|^2.$$

For $q>1$, we define the moment sum
\beq\label{M_q}
M_q(\mu_\lambda)=\sum_{\xi\in\Z^d}\mu_\lambda(\xi)^q.
\eeq
A fractal scaling law, in the semiclassical limit $\lambda\to\infty$, is a power law
\beq
M_q(\mu_\lambda)\sim N_\lambda^{(1-q)D_q},
\eeq
where $N_\lambda$ denotes the number of $O(1)$-contributions in \eqref{M_q}, as $\lambda\to\infty$, and $D_q$ denotes the fractal exponent.

\begin{figure}
\includegraphics[scale=0.5]{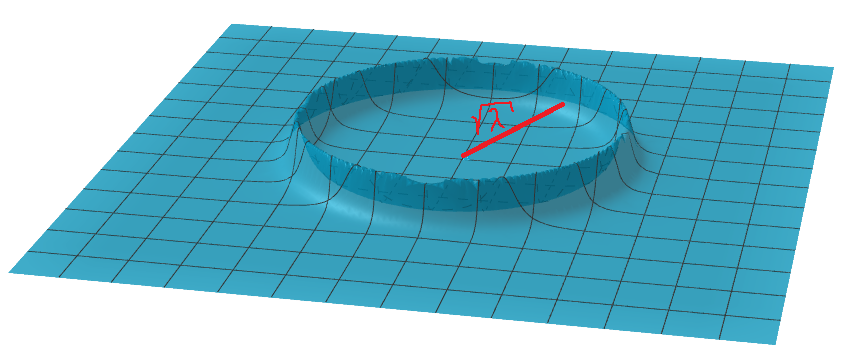}
\caption{The measure $\mu_\lambda$ is concentrated on lattice points which lie inside a thin annulus of central radius $\sqrt{\lambda}$. The width of this annulus grows with $\lambda$ on a logarithmic scale. The number of lattice points inside the annulus is subject to subtle fluctuations.}
\label{fig1}
\end{figure}

We note that, in the case of the torus, the mass of the probability measure $\mu_\lambda$ is concentrated on lattice points which lie inside a thin annulus of central radius $\sqrt{\lambda}$ and whose width depends on the spectral parameter $\lambda$ (cf. Figure \ref{fig1}). In the semiclassical limit, as $\lambda\to\infty$, the number of $O(1)$-contributions will grow slowly (in fact on a logarithmic scale) with $\lambda$. However, this number fluctuates a lot, as the number of lattice points in thin annuli is subject to strong fluctuations (this is due to the scale of the width being of much lower order than the error term in the Gauss circle law). In order to compute $N_\lambda$, as a function of $\lambda$ one must perform a spectral average. This is the first challenge, from a mathematical point of view, to be able to prove a fractal scaling law. As we will see below, for particularly simple choices of potential, where the measure $\mu_\lambda$ takes a simple and explicit form, it is possible to perform this calculation. For generic potentials, it is expected to be a much more challenging task.

In order to compute the fractal exponent associated with a sequence of eigenfunctions, we introduce the Renyi entropy of the measure $\mu_\lambda$:
\beq
H_q(\mu_\lambda)=\frac{1}{1-q}\log M_q(\mu_\lambda), \quad q>1.
\eeq
The Renyi entropy may be thought of as a generalization of the Shannon entropy which is familiar from information theory, in the sense that the latter is recovered in the limit, as $q\to 1$:
$$
\lim_{q\to 1} H_q(\mu_\lambda)=-\left[\frac{d}{dq}\log M_q(\mu_\lambda)\right]_{q=1}=-\sum_{\xi\in\Z^d}\mu_\lambda(\xi)\log \mu_\lambda(\xi).
$$

Provided one can obtain an asymptotic for the Renyi entropy in the limit, $\lambda\to\infty$, possibly by restricting oneself to a subsequence of eigenvalues, and on tackles the problem of determining the scaling parameter (by averaging out the fluctuations mentioned above), then one might hope to be able to compute the fractal exponent $D_q$ for $q>1$.

For a generic choice of the potential $V$ this problem can be very hard. To give an idea of the challenges involved: if one picks a potential modelling a disordered system in a scaling regime that corresponds to the thermodynamical limit (say taking a large torus and scaling back to the standard torus), the occurrence of multifractal scaling appears to be related to the onset of a phase transition between localized and delocalized regimes (in $d\geq 3$).

For a simple choice of potential, however, which allows for explicit expressions of the eigenfunctions, and, thus, the measure $\mu_\lambda$ it is possible to overcome these challenges.

\section{Multifractality for an arithmetic \u{S}eba billiard}
In a 1990 paper \cite{Se90} Petr \u{S}eba introduced rectangular billiards with a Dirac delta potential placed in the interior as a toy model for more complicated pseudo-integrable billiards whose dynamics is in some sense intermediate between integrable and chaotic dynamics. In this section we will consider a slightly modified version of this billiard, namely a square torus with a delta potential. We will refer to this as an arithmetic \u{S}eba billiard, because the Laplace spectrum is of arithmetic nature. It is given, up to a factor, by integers representable as a sum of two squares:
$$\sigma(-\Delta_{\T^2})=\{n=x^2+y^2 \mid (x,y)\in\Z^2\}$$
We note that the Laplace eigenvalues have multiplicities which are given by the arithmetic function
\beq
r_2(n)=\#\{(x,y)\in\Z^2 \mid n=x^2+y^2\}
\eeq
which counts the number of lattice points on the circle of radius $\sqrt{n}$.

Employing self-adjoint extension theory one can show that the spectrum of the \u{S}eba billiard consists of two types of eigenvalues. There are old Laplace eigenvalues, with multiplicity reduced by $1$, which correspond to co-dimension $1$ subspaces of eigenfunctions which vanish at the position of the potential. There are also new eigenvalues, with multiplicity $1$, corresponding to new eigenfunctions which feel the potential. These new eigenvalues interlace with the Laplace eigenvalues.
 
Moreover, self-adjoint extension theory yields explicit formulae for these new eigenfunctions which in turn give rise to an explicit expression for the Fourier coefficients and, hence, the measure $\mu_\lambda$: 
$$\mu_\lambda(\xi)=\frac{(|\xi|^2-\lambda)^{-2}}{\sum_{\xi'\in\Z^2}(|\xi'|^2-\lambda)^{-2}}$$
Moreover, we note that $\lambda\notin\sigma(-\Delta)$, because of the interlacing property of the new eigenvalues.

The moment sums associated with the measure $\mu_\lambda$, for a new eigenvalue $\lambda$, are of the form
\beq
M_q(\mu_\lambda)=\frac{\zeta_\lambda(2q)}{\zeta_\lambda(2)^q},
\eeq
where we introduce the shifted zeta function
\beq
\zeta_\lambda(s)=\sum_{n\geq0}\frac{r_2(n)}{|n-\lambda|^s}, \quad \Re s>1.
\eeq

\subsection{Weak coupling: a monofractal regime}
It is instructive to look at the physically trivial case of weak coupling (fixed self-adjoint extension).
In this regime, $\lambda$ is typically close to a neighbouring Laplace eigenvalue $m$ (cf. \cite{RU12}). Let us denote by $\Delta_j$ the distance between a new eigenvalue and the nearest Laplace eigenvalue. 

For a given $x\gg 1$, we define the mean distance up to threshold $x$ as
\beq
\bra\Delta_j\ket_x=\frac{1}{\#\{\lambda_k\leq x\}}\sum_{\lambda_k\leq x}\Delta_k.
\eeq
In the case of the square torus, we have $\bra\Delta\ket_x=O((\log x)^{-1/2})$ (which is a special case of a more general estimate derived in \cite{RU12}), where we note that in this case the average spacing of the Laplace eigenvalues is of order $\sqrt{\log x}$ due to the multiplicities in the Laplace spectrum.

Thus, only one term (or one circle in the lattice with radius $\sqrt{m}$) contributes. The sum scales as follows along the subsequence of typical eigenvalues:
$$M_q(\mu_\lambda)=\frac{\zeta_\lambda(2q)}{\zeta_\lambda(2)^q}\sim \frac{r_2(m)|m-\lambda|^{-2q}}{(r_2(m)|m-\lambda|^{-2})^q}=r_2(m)^{1-q}$$

The number of terms which contribute is simply the number of lattice points on the circle $|\xi|^2=m$. 

The Renyi entropy has asymptotics
$$H_q(\mu_\lambda)\sim\frac{1}{1-q}\log(r_2(m)^{1-q})=\log r_2(m).$$

It can be shown that for a full-density subsequence of Laplace eigenvalues we have for any $m$ in this subsequence 
$$r_2(m)=(\log n)^{\tfrac{1}{2}\log 2+o(1)}, \quad m\to+\infty.$$ 
Hence, $$N_\lambda=(\log m)^{\tfrac{1}{2}\log 2}$$ which is known as the normal order of $r_2$.

We note that the fluctuations of the arithmetic function $r_2(n)$ are very subtle. It is a classical theorem of Landau from 1907 that the number of integers less or equal than $x$ grows like $cx/\sqrt{\log x}$ which implies that on average the multiplicities are of order $\sqrt{\log x}$. The smaller exponent $\tfrac{1}{2}\log 2$ arises along a typical (as in full density) subsequence, because there is a very sparse subsequence, where $r_2(n)$ grows much faster (of order $n^{o(1)}$ for some slowly decaying exponent function). Moreover, there are also sparse subsequences, where $r_2$ remains bounded.
 
From the Renyi entropy one can now readily obtain the fractal exponent 
$$D_q=\lim_{\lambda\to\infty}\frac{H_q(\mu_\lambda)}{\log N_\lambda}=1.$$

In particular, we note that the fractal exponent does not vary with $q$, because, due to the weakness of coupling strength, only the nearest circle contributes. 
 
\subsection{Strong coupling: a multifractal regime}
The physically interesting regime requires a renormalization of the extension parameter in the semiclassical limit. This allows to consider stronger coupling strength. We can measure the strentgh of the perturbation by computing the mean distance between old and new eigenvalues. For a suitable renormalization one obtains
$$\bra\Delta_j\ket=(\log x)^{\alpha+o(1)}, \quad \alpha\in(-1/2,1/2],$$
where the exponent $\alpha$ is a measure of the strentgh of the perturbation. 

Because in such regimes the new eigenvalues lie farther away from the neighbouring Laplace eigenvalues (on the scale of the mean spacing of the eigenvalues), many more circles contribute. In fact, all lattice points in a thin annulus of central radius $\sqrt{\lambda}$ must be taken into account. 

We have the following theorem, proven jointly with Keating in \cite{KeU21}, which computes the fractal exponents associated with a full density subsequence of new eigenvalues in a strong coupling regime. For a range of exponents $q$ which depends on the coupling strength $\alpha$ associated with the subsequence we derive an explicit formula for the fractal exponent which shows how it varies with $q$, thereby proving multifractality.
\begin{thm}
Let $\Lambda$ be a sequence of new eigenvalues in a strong coupling regime such that $\alpha(\Lambda)\in(\frac{1}{4},\tfrac{1}{2})$. There exists a full-density subsequence $\Lambda'\subset\Lambda$ such that for any $q$ in the range 
$$\frac{1-\log 2}{2-4\alpha}<q\leq \frac{1}{2-4\alpha}$$ we have the following formula for the fractal exponents associated with the sequence $\Lambda'$:
\beq
D_q(\Lambda')=\frac{1}{2\alpha}(1-\frac{1}{2q})\log 2
\eeq
\end{thm}

\subsection{The ground state regime}
Instead of studying a high frequency regime, where $\lambda\to\infty$, one might as well consider a low frequency regime, where $\lambda\to 0$. In this regime there is no relationship expected between the intermediate type of dynamics and the occurrence of multifractality. Rather, multifractality in such regimes is expected to occur for a much wider class of systems.

However, in the case of \u{S}eba billiards there is a very interesting link with Epstein's zeta function associated with quadratic forms. This link occurs for general tori not just arithmetic ones. We introduce the following modified version of the shifted zeta function above:
\beq
\zeta_\lambda^*(s)=\sum_{n\in\cN}\frac{r_Q(n)}{|n-\lambda|^s}, \quad \Re s>1,
\eeq
where $\cN$ denotes the Laplace spectrum on a general unimodular rectangular tori, given by the set of values taken by the quadratic form $Q(x,y)=a^2x^2+a^{-2}y^2$, $(x,y)^2$, and $a>0$. Moreover, $r_Q$ denotes the representation number of $Q$.

We introduce the modified moment sums $M_q^*(\lambda)=\zeta_\lambda^*(2q)$ for $q>1$. Note that we need to remove the first term, as this blows up in the limit $\lambda\to 0$. We are, thus, interested in the fluctuations around this blow-up term which motivates the study of the modified moment sums.

For $q>1$ we define the fractal exponents
\beq
D_q^*=\frac{d_q^*-qd_1^*}{q-1}, \quad d_q^*=\lim_{\lambda\to0}\zeta_\lambda^*(2q)=\zeta_Q(2q),
\eeq
where we denote Epstein's zeta function associated with the quadratic form $Q$ as
$$\zeta_Q=\sum_{(m,n)\in\Z^2\setminus\{0\}}Q(m,n)^{-s}, \quad \Re s>1.$$
We also note that we have the following functional equation
$$\zeta_Q(1-s)=\varphi_Q(s)\zeta_Q(s)$$
where $\varphi_Q$ denotes a certain meromorphic function associated with $Q$.

The first predicition of a symmetry relations for the fractal exponents of multifractal systems is due to Mirlin, Fyodorov, Mildenberger and Evers for the case of the Anderson model in\cite{MirFyMilEv06}. The following symmetry relation was proved in \cite{KeU21}.
\begin{thm}
The fractal exponent $D_q^*$ admits an analytic continuation to the full complex plane. It satisfies the following symmetry relation with respect to the critical point $q=1/4$:
\beq
D_{1/2-q}^*=\frac{1-q}{1/2+q}\left(D_q^*+\frac{\log\varphi_Q(2q)+(2q-1/2)\log\zeta_Q(2)}{1-q}\right)
\eeq
\end{thm}

\section{Outlook}
Multifractal scaling is an important property of quantum systems which are intermediate between two physical regimes, and many important systems such as the Anderson model and pseudo-integrable quantum billiards fall into this category. However, understanding the rigorous mathematical underpinning of multifractality goes far beyond the study of intermediate quantum systems. In fact, multifractal scaling appears to be related to deep and important mathematical problems in a number of models in mathematical physics. Another highly interesting type of models are nonlinear partial differential equations such as the Euler and Navier-Stokes equations which model the dynamics of incompressible fluids. In this case, it turns out that the occurrence of multifractal scaling is related to the deep and difficult question of the regularity or blow-up of solutions to these nonlinear PDE which is the subject of a forthcoming article \cite{U23}.

\end{document}